\documentclass[10pt,a4paper]{article}
\usepackage[margin=1.0in]{geometry}

\usepackage[ruled]{algorithm2e}
\usepackage{amssymb}
\usepackage{amsmath}
\usepackage{array}
\usepackage{mdwmath}
\usepackage{mdwtab}
\usepackage{url}
\usepackage{graphicx}
\usepackage[normalem]{ulem}
\usepackage{paralist}
\usepackage{placeins}
\providecommand{\keywords}[1]{\textbf{\textit{Index terms---}} #1}
\date{\vspace{-5ex}}
\begin{document}
	
	\title{When to Build Quantum Software?}
	\author{
		Janardan Misra, Vikrant Kaulgud, Rupesh Kaslay\footnote{Contributed while with Accenture Labs, Bangalore, India}, Sanjay Podder \and
		Accenture Labs, Bangalore, India\and
		\{\sf{janardan.misra,vikrant.kaulgud,sanjay.podder}\}{\sf @accenture.com}
	}

\maketitle

\begin{abstract}
  Despite ongoing advancements in quantum computing, businesses are still faced with the problem to decide if they would benefit from investing into this novel technology for building a business critical application. This uncertainty is not only owing to the limitations in the current state of the technology but also due to the gap between the level at which business applications are analyzed (e.g., using high level semi-formal languages) and the level at which quantum computing related information is currently available (e.g., formally specified computational problems, their algorithmic solutions with computational complexity theoretic analysis) to make informed decisions. To fill the discourse gap, in this paper, we present design of an interactive advisor, which augments users while deciding to invest into quantum software development as a plausible future option in their application context. Towards that we apply business process modeling and natural language similarity analysis using text-embeddings to associated business context with computational problems and formulate constraints in terms of quantum speedup and resource requirements to select software development platforms.
\end{abstract}


\maketitle

\section{Introduction}

As the promise of quantum computing (QC) from theory to practice is increasingly being considered as realistic in near future~\cite{8,corcoles2019challenges}, businesses across spectrum have started looking forward to its adoption in their application architecture and development life-cycles.
 
However, as the current-state of QC technology and its limitations exist (in particular, at the hardware  stack including decoherence, high error-rates, and low qubit counts)~\cite{aaronson2008limits}, most of the demonstrated applications are limited to specific set of relatively low-level algorithmic challenges and operate at prototype level. Examples include prime factorization (for cryptographic applications), adiabatic optimization (for constraint satisfaction problems), molecular simulations (for drug discovery) etc.~\cite{qalgo,preskill2018quantum}. 

Furthermore, despite emergence of many quantum development platforms and programming languages~\cite{qcomputingReview,larose2019overview,bichsel2020silq}, developing quantum software remains a cost intensive process since it requires skills and resources not typically acquired through traditional software development practices. 

So, a challenge businesses are facing is how to decide whether their applications will truly benefit from investment into QC early on or should they wait before it is mature enough to become a commercially viable option. Both these options have trade-offs to deal with. Early investments while having a potential to bring competitive business edge, however, involve strategic uncertainties since it is not clear exactly how long we need to wait before reliable and scalable quantum computers are ready for commercial application development. On the other hand, if businesses wait long enough, they may miss opportunity and later may have to struggle to pick up fast enough with possible cost escalations. Therefore, in order to balance these trade-offs well, a pragmatic approach is required to establish clear business case as per the technology road map of the application and invest strategically. 

However, there exist gap between the level at which business applications are analyzed and the level at which QC related information is currently available to make informed decisions. This gap exists because users having business knowledge and high-level functional requirements (e.g., business analysts, requirements analysis) have been traditionally expressing and analyzing business requirements and associated details in terms of use-cases, business process models, etc. Whereas most of the information currently discussed about QC is at the level of computational problems and algorithms including details on computational complexity theoretic related aspects of quantum algorithms as well as corresponding classical solutions (see~\cite{qalgo} for a high-level overview). 
	
This gap is not easy to reliably fill even for those experienced in design and development of classical software (e.g., technical architects, software developers) because design, development, and analysis of the quantum software is fundamentally different from classical software development practices prevalent today~\cite{qse}. Owing to this gap in the level of discourse, it is difficult for business users to take informed decisions without consulting QC specialists. In particular, there does not exist any decision support tool to augment the users by bridging the gap. To address this limitation, in this work, we present an approach to augment users while deciding to invest into quantum software development as a plausible future option in their application context. 
	 
	

\section{Approach}
At a high level, we adopt the following approach to meet these challenges: Designing an interactive advisor, which we will refer henceforth as \textit{Quantum Business Requirements Analyzer} (QBRA) to fill the discourse gap. QAB starts at the level of business applications and gradually guides its users to the level where QC related decision can be taken. It maintains an evolving list of applications and applies Natural Language Processing (NLP) techniques~\cite{spl3} to address the challenge of determining if an application is suitable for QC. 

\begin{figure*}
	\includegraphics[width=\textwidth]{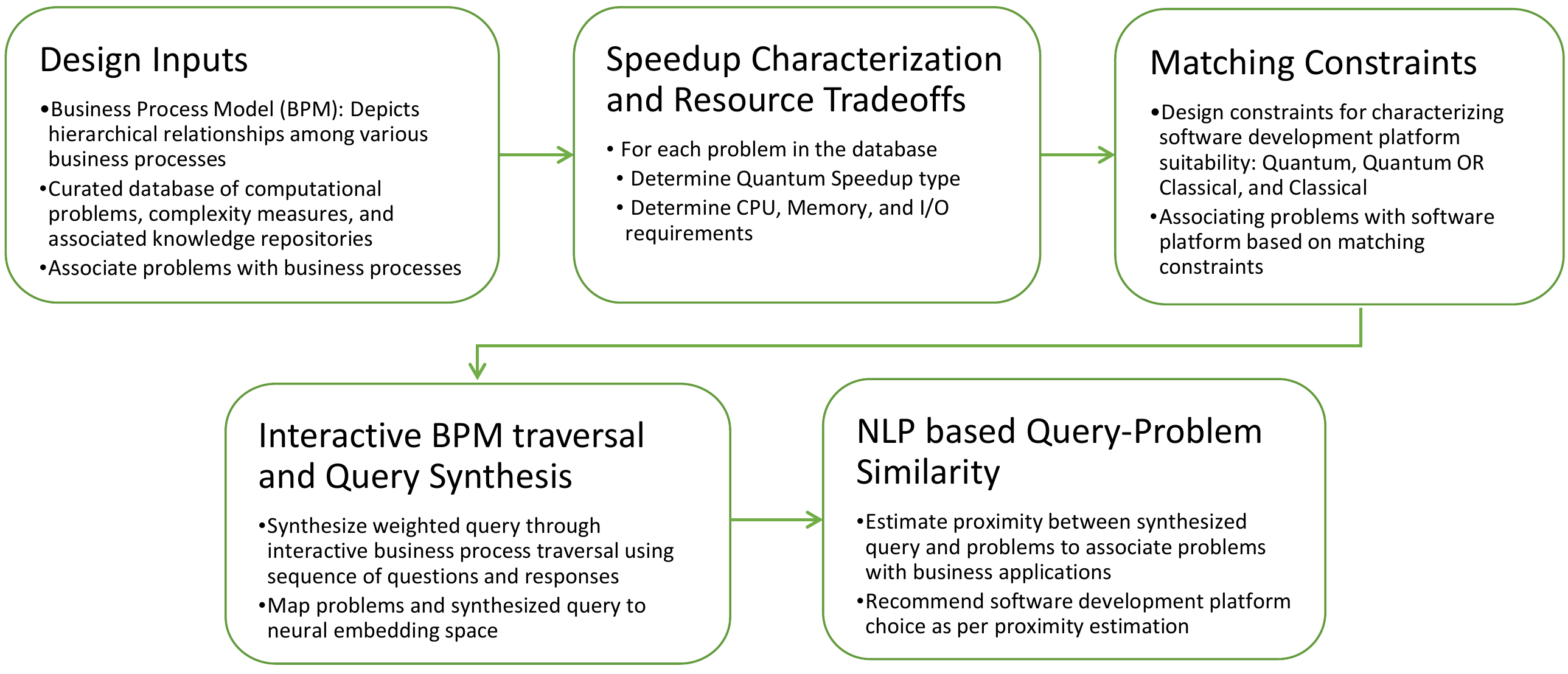}
	\caption{High Level Process Flow of Quantum Business Requirements Analyzer}
	\label{processflow}
\end{figure*}

Figure~\ref{processflow} depicts the high-level process flow of QBRA at design and execution stages. Forthcoming sections provide details on various subsections.

{\bf Interactive Business Process Traversal:}\label{guided} When a user only has a high-level view of the application, e.g., its business and functional objectives, a survey-like structured approach can be used to capture the user’s intent and guide him/her interactively towards the space where relevant QC related technical details are available. In other words, QBRA guides its users to make choices by navigating through a business process model (BPM)~\cite{bpmn} in a question and answer mode. Such BPM based guided traversal captures both high-level business objective(s) as well as guides the user to level of technical objective(s) in order to inform whether QC or classical methods should be used for developing an application to meet those objectives.  

BPM navigation can be designed by generating questions using templates or by associating curated questions with each level and nodes in the BPM hierarchy or designing a natural dialogue system (e.g., chatbot)~\cite[Chap 26]{spl3} using BPM as underlying knowledge graph.

{\bf Templatized Questionnaire} At each level of the BPM hierarchy, QAB agent generates a question, which can be formulated using the below template: $$Q_{cur}\equiv Concatenate(Str_1, Sel_{pre}, Str_2, Op_{cur})$$ where $cur = \textit{current level (initialized to 1)}$,  $pre = cur - 1$, $Sel_{pre}\equiv$ is the user selection at previous level, $Op_{cur}$ is the set of child nodes of $Sel_{pre}$ at current level, $Str_1 \equiv$ ``Which of the following under”, and $Str_2 \equiv$ ``are related to the intended application?".	
For example, according to this template, a question under financial services could look like: {\it ``Which of the following options under Financial Services are related to the intended application? 1. Capital Market 2. Banking 3. Insurance"}

For example, lets us illustrate curated dialogue flow by assuming a business analyst as its user, who is familiar with the domain of financial services and needs to decide whether QC will be suitable for an application related to `analysis of customer portfolios'. In this case, QBRA will guide user to make appropriate choices using following sequence of interactions: {\bf Q1} Please select your industry? \textit{User Selection}: Financial Services. $\rightarrow$ {\bf Q2} Please select type of financial services for your application? \textit{User Selection}: Capital Markets. $\rightarrow$ {\bf Q3} Please select type of resources to be managed? \textit{User Selection}: High Performance Assets. $\rightarrow$ {\bf Q4} Please select type of high performance assets your application is going to manage? \textit{User Selection}: Portfolios. $\rightarrow$ {\bf Q5} What is the key functionality of the application in relation to performance measurement of funds? \textit{User Selection}: Input Data Set-up. $\rightarrow$ {\bf Q6} Is input data set-up related to portfolio construction? \textit{User Selection}: Yes. $\rightarrow$ {\bf Q7} Does it need to optimize portfolio selection? \textit{User Selection}: Yes.
 
At this point QBRA narrows down to `portfolio optimization' as intended technical objective and is followed by the dialogue below which is aimed towards eliciting low level functional objectives of the application under `portfolio optimization'.
\begin{description} 
\item[Q8] Please select problem scenario most closely aligned with the application:
\begin{itemize}
	\item {\sf Scenario 1}: Find the best stocks to incorporate into a portfolio for a given price.
	\item {\sf Scenario 2}: Find portfolio constitution having minimum risks in long term.
	\item {\sf Scenario 3}: Find stocks which will collectively give maximum profit in medium term.
\end{itemize}
\end{description}
Based upon the selection of the problem scenario, QBRA will make recommendations as discussed in Section~\ref{reco}.

{\bf Problem Database:} In the above sequence of user interactions with QBRA, initial questions [Q1]--[Q7] were generated by navigating the BPM of the financial-services, whereas last three problem scenarios were extracted from the QBRA problem database (QBRApDB). QBRApDB consists of curated list of computational problems, which are mapped to the leaf nodes in BPM, i.e.,  for each of the nodes in the lowest level of business process hierarchy a set of computational problems appearing in QBRApDB are curatively linked. 

Mapping between computational problems in QBRApDB and leaf nodes in the BPM forms basis of recommendations made by QBRA to it operating environment. It is discussed next:
 
Each computational problem $p\in QBRApDB$ is associated with following fields: \[\left(def(p),cc_q(p), cc_c(p), KD(p)\right)\]
\begin{itemize}
	\item $def(p)$ is computational formulation of the problem $p$. 
	\item $cc_q(p)$ be the computational complexity of the most efficient known quantum algorithm for solving $p$.
	\item $cc_c(p)$ be the computational complexity of the most efficient known classical algorithm for solving $p$.
	\item $KD(p)$: A corpus of textual information from accessible knowledge sources associated with $p$. 
\end{itemize}
In above format, QBRA problem database contains problems for which either quantum or classical solutions are known. Examples include
\begin{itemize}
	\item Pre-identified QC use-cases including problems listed in `Quantum Algorithm Zoo'~\cite{qalgozoo}. 
	\item Quadratic Unconstrained Binary Optimization (QUBO) Problems~\cite{6}.
	\item Problems related to Simulation of Physical Systems.
	\item Optimization problems in machine learning.
\end{itemize}

For each of the nodes in the lowest level of business process hierarchy, a set of computational problems appearing in the QBRA problem database are manually linked, which form the basis of its recommendations as discussed in Section~\ref{reco}. 

{\bf Weighted Query Generation:} Using BPM navigation, QBRA builds a query by interactively guiding its user to elicit relevant details about the business application of interest. 

Initialization: $Query(final)$ = $Query_{app}(l_0)=\emptyset$, where $l_0$ indicates starting level of the business process navigation, e.g., name of the industry. $\emptyset$ (empty query) indicates that no relevant information is available at this level to make a decision.

With every selection by the user, QBRA   augments the query such that after navigating the $r^{th}$ level: $Query_{app}(l_r )$ = $Query_{app}(l_{r-1}) + Ch_r$, where $Ch_r$  is the concatenation of [QBRA Interaction Phrase] and choice made at the $r^{th}$ level by the user. QBRA interaction phrase is discussed earlier while discussing design choices for QBRA interactive traversal of the BPM. [] indicates that QBRA interaction phrase is optional. Finally, $Query_{app}(final)$ =$Query_{app}(l_r)$.

After making selection at each level (except at the leaf level), the user makes one of the two choices: either continue to explore BMP to the next level or close the navigation process. On closure of the guided navigation process, QBRA   fires the query $Query_{app}(final)$ against its problem database as will be further discussed in the next section.
 
For example, in reference to the BPM interactions before, QBRA will build its query as follows: $Query_{app}(l_1)$: [Q1] + ``financial services”  $\ldots$ $Query_{app}(final)$ = $Query_{app}(l_{9})$: [Q1] + ``financial services” + [Q2] + ``capital markets” + ... + [Q9] + [Scenario 1]. 

{\sf Estimating information theoretic weights (or relative relevance) of query terms:} In order to improve accuracy of mapping, let a term denote phrases consisting of one or more words describing a node in the BPM hierarchy. Let $N$ = number of nodes in the BPM estimated as $\Sigma_{l\in BPM}\mathit(Number\ of\ nodes\ at\ l^{th}\ level)$. For each of the terms $x$ in the BPM, further estimate $\Delta(x)$ = Number of nodes in the subtree rooted at $x$. In terms of these, relative relevance or weight of a term $x$ is estimated as $$\alpha(x)=log\left(\frac{N}{\Delta(x)+1}\right)$$

{\bf Mapping Problem and Query onto Neural Embedding Space:} All the information associated with each problem in QBRApDB is succinctly represented as a  neural embedding (e.g., as Glove vectors~\cite{glove}). Let $em(p)$ represent the neural embedding of the problem $p$ together with its coupled knowledge sources. Next, user query $Q_{app}(final)$ describing the intended application is also mapped to the same embedding space. Let $em(Query_{app}(final))$ be the embedding, which is a weighted average of the embeddings of the terms extracted from the BPM navigation with weights being the relative relevance estimated before:
\[em(Query_{app}(final)) =  \frac{1}{s}\sum_{x \in Query_{app}(final)}(\alpha(x)\times em(x))\]
where $s$ is the number of terms in the query $Query_{app}(final)$.

{\bf Matching:} Next, a vector matching method (e.g., cosine similarity~\cite{cosine}) is applied to determine which of the existing problems in the QBRApDB are semantically close to the generated query.

$match(Query_{app}(final), QBRApDB)$\[
= 
\begin{cases} 
p & \text{if } \left((\delta \leq d_p)\ AND\ (\forall_{q\neq p \in QBRApDB} (d_q \leq d_p))\right) \\
\emptyset & \text{otherwise} 
\end{cases}
\] where $d_p = cosine(em(p), em(Query_{app}(final)))$ and $d_q$ = $cosine(em(q), em(Query_{app}(final)))$ and $\delta \in [0, 1]$ is the threshold of similarity so that only those problems with similarity more than $\delta$ are considered for making recommendations. 

If the method $match()$ returns non-empty set of problems, the  recommendation corresponding to the closest match is presented to the operating environment as QBRA output (ref. Section~\ref{reco}). Else if there is no match i.e., match returns empty set, QBRA makes recommendation as specified by its design environment. For example, recommendation to consult an algorithm design expert to confirm if application would require solving computationally hard problems (e.g., combinatorial optimization problems)

\section{Speedup Characterization and Computational Resource Requirements}

{\bf Quantum Speedup Estimation:} For each problem $p \in QBRApDB$, in terms of $cc_c(p)$ and $cc_q(p)$, computational speedup of the fastest known quantum algorithm over the classical one is estimated as \[speedup(p,cc_{c}(p),cc_q(p)) = cc_{c}(inverse(cc_q(p)))\] 
where $inverse(.)$ returns functional inverse of its argument function. For example, let $cc_{q}(p)=\mathcal{O}(\log{}n)$ and $cc_{c}(p)=\mathcal{O}(n^3)$, where $n$ is the instance size. 
\begin{align*}
speedup(p,cc_{c}(p),cc_q(p)) &= \mathcal{O}((inverse(\log{}n))^3) \\ 
 &= \mathcal{O}\left(\left(2^n\right)^3\right) = \mathcal{O}(2^{3n})\\ 
 &= \mathcal{O}(2^{n})\ \textit(exponential)
\end{align*} 
{\bf Speedup Type:} For simplicity, we divide speedup in three broad categories
\begin{itemize}
	\item {\bf EXP+} If $speedup()$ is an exponential or slower function. For example,  $\mathcal{O}(2^{cc_q(p)})$ (exponential) or $\mathcal{O}(2^{2^{cc_q(p)}})$ (super exponential).
	\item {\bf POLY+} If $speedup()$ is slower than a polynomial function but faster than exponential. For example,
	$\mathcal{O}(cc_{cl}(p)^{\log{}(cc_q(p))})$ (superpolynomial).
	\item {\bf POLY-} If $speedup()$ is is an polynomial or faster than polynomial function. For example,  $\mathcal{O}(cc_q(p)^k)$ (polynomial) or $\mathcal{O}(\log{}(cc_q(p)))$ (logarithmic).
\end{itemize} 

{\bf CPU-Memory-I/O Requirements:} QBRA design environment determines CPU-memory trade-offs for the most efficient classical algorithm of the problem $p$ on a classical software platform using its simulated runs.
\begin{itemize}
	\item {\bf CPU Bound}: Problem $p$ is considered as CPU bound if majority of its time is spent on performing arithmetic calculations. For example generating number series.
	\item {\bf I/O Bound} Problem $p$ is considered as I/O bound if majority of its time is spent on performing read/write operations on external storage (secondary storage or network storage). For example Information Retrieval tasks.
	\item {\bf Memory Bound} Problem $p$ is considered as Memory bound if during it executions, majority of its time is spent on manipulating data on its local memory (RAM). For example: Matrix operations.
\end{itemize}

\section{Matching Applications with Software Development Platform}\label{reco}
Based upon the above, for each of the problems in its  database, QBRA maintains a curated recommendation involving type of software development environment likely to be suitable to build the application and links to the accessible knowledge sources (e.g., case-studies). QBRA makes recommendations of the below types:
	
\subsection*{\bf Case 1: Hybrid Quantum Architecture} A problem $p\in QBRApDB$ is classified as one suitable for QC (with mixed architecture) under NISQ (Noisy Intermediate Scale Quantum)~\cite{preskill2018quantum} era if it satisfies following constraints:\[\left(C_1\ AND \ C_2\right)\ OR\ \left(C_1\ AND\ C_3\ AND\ C_4\right)\] where $C_1$: Speedup Type of $p$ is EXP+, $C_2$: $p$ is CPU bound, $C_3$: $p$ is MEMORY bound, and $C_4$: Intermediate memory consumption of $p$ is high. Informally, if there exists a quantum algorithm having exponential speedup over best known classical algorithm and problem is either compute intensive or memory intensive, quantum software development using system-of-system approach involving both quantum and classical components could be a preferred option especially for future readiness perspective. 
	
If an application is associated with such a problem $p$, QBRA recommends ``Quantum Computing (with hybrid architecture involving Classical Components) is a plausible choice under NISQ." Adding further that problem $p$ underlying the application has no known efficient classical algorithm for instances of practical significance. Today’s QC technology may help in simulating a solution and a hybrid architecture with both quantum as well as classical components is a recommended direction to invest in near term.
	
\emph{Resource Estimates}: Additional insights may be of help in deriving a rough estimate of quantum resources required to solve the selected problem. This can be achieved either by curating details from the literature or if quantum prototype is available, using tools like \textit{Trace simulator} bundled with Microsoft's Quantum Development Kit~\cite{qdkit}, which provides estimates for counts of CNOT gate, measurements, rotations, etc. 
	
\subsubsection*{Example} Suppose an organization from pharmaceutical industry is looking to build an application to help during drug discovery process, QBRA would recommend that such application is better suited for QC because the problem of simulating properties of (drug) molecules is hard  for classical systems (especially when high accuracy calculations are required), and Quantum programs have inherent advantage in such quantum state simulations, especially in the light of recent advancements in the field of quantum computational chemistry~\cite{grimsley2019adaptive,qchemistry1,qchemistry2}. Additionally, QBRA would add that for practical purposes, as per current state-of-the-art, few million physical qubits are necessary for practical scenarios.  
	
\subsection*{\bf Case 2: Classical or Quantum}  A problem $p\in QBRApDB$ is classified as one suitable for classical software development or QC (with mixed architecture) if it satisfies following constraint:\[((C_5\ OR\ C_6)\ AND\ C_7)\ AND\ (C_2\ OR\ (C_3\ AND\ C_4))\] where $C_5$: Speedup Type of $p$ is POLY+, $C_6$: Speedup Type of $p$ is POLY-, and $C_7$: $cc_{cl}(p)$ is faster than exponential. Informally, if the speed-up of best known quantum algorithm for $p$ over classical algorithms is not very high and classical algorithm for solving $p$ is feasible and $p$ is a CPU bound problem, then it can be developed either using classical or quantum computing platforms.
	
If an application is associated with such a problem $p$, QBRA recommends ``Both Classical as well as Quantum Computing (with hybrid architecture) are plausible choices. Underlying computational problem for this application can be solved approximately using classical methods in polynomial time. For more accurate solutions or for computational speedups on larger instance sizes, a quantum processing unit with hybrid architecture may be a better choice." 
	
		
\subsubsection*{Example}  For the illustrative use-case of the Guided Mode presented in Section~\ref{guided}, if user selects {\it problem scenario 1}, QBRA will recommend that both CC and QC are plausible choices~\cite{portfolio}. 
	
\subsection*{\bf Case 3: Classical} A problem $p\in QBRApDB$ is classified as one suitable for classical computing if it satisfies following constraint:\[(C_6\ AND\ C_8)\ AND\ (C_9\ OR\ (C_3\ AND\ C_{10}))\] where $C_8$: $cc_{cl}(p)$ is polynomial or faster, $C_9$: $p$ is I/O bound, and $C_{10}$: Input or output volume of $p$ is high. Informally, if speed-up of best-known quantum algorithm for $p$ as compared to classical algorithms is not much and best known classical algorithm for $p$ is efficient and $p$ is a memory bound problem having high intermediate memory requirements, building software to execute on classical platform is sufficient.
	
If an application is associated with such a problem $p$, QBRA recommends ``Classical Computing is Sufficient. The computational problem underlying this application has well-known classical algorithm(s) that provide exact solutions.” Furthermore, if a use-case study exists in the QBRA knowledge base, it would add something like ``Here's an example of how it was solved in [\textit{similar business context}].”

\subsubsection*{Example}  Suppose a startup in agriculture industry is looking to help farmers by designing a mobile  recommendation app to suggest which combination of crops to grow and how much should be the quantity of each crop  to maximize the overall yield having met constraints on available farming area, seed costs, and expected yields from each crop. Such farm planning problems can be modeled as Linear Programming (LP)~\cite{agriLP}. Therefore, QBRA would recommend it to be case of CC and point to recent works like~\cite{2} to efficiently solve it. 

\section{Tool Design and Validation}
\begin{figure*}
	\centering
	\includegraphics[width=\textwidth]{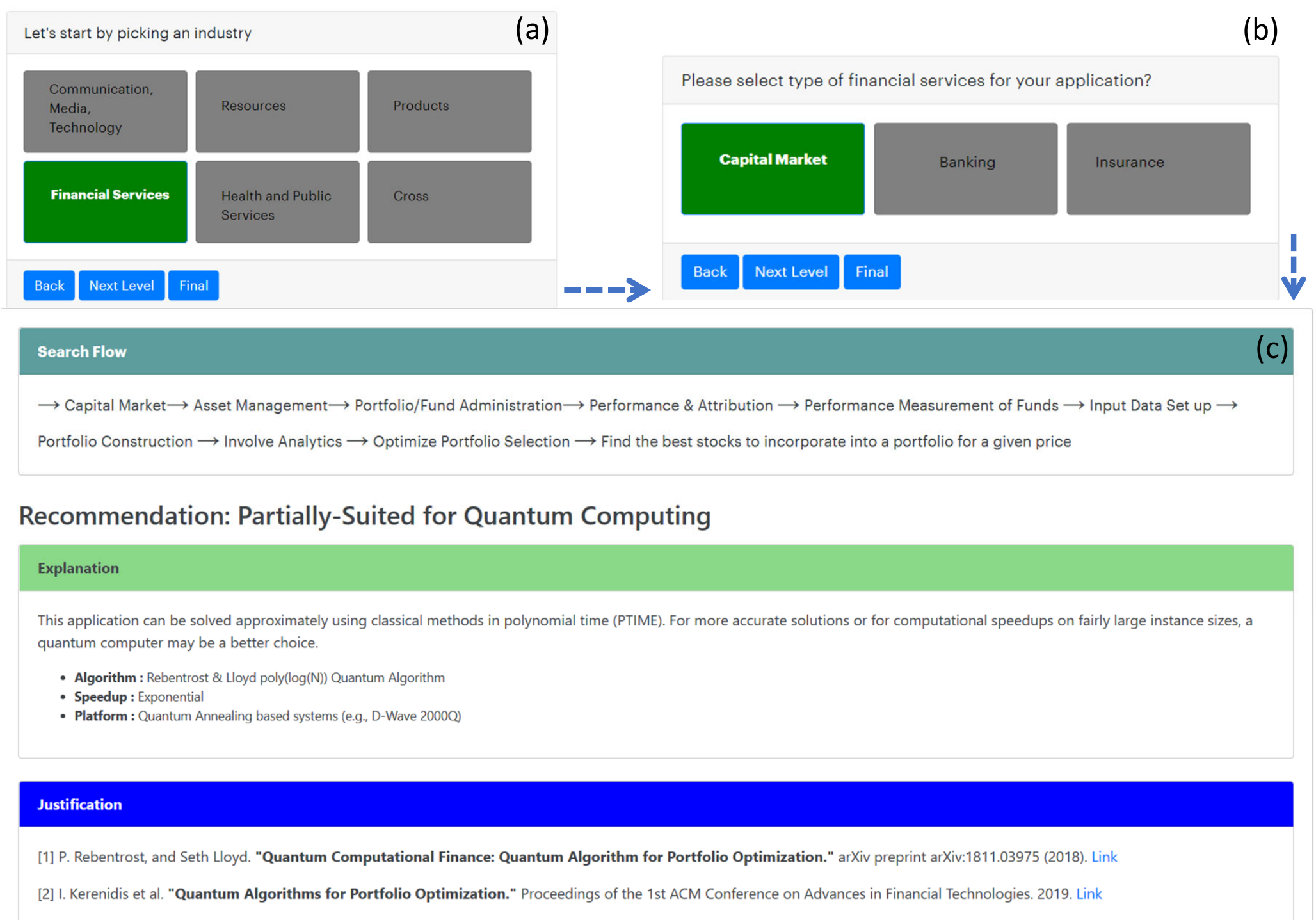}
	\caption{QBRA Prototype tool UI Flow (\textit{selected screens}): (a) QBRA starts with asking its user to select the industry where its application is intended to work, (b) QBRA thereafter narrows down the application focus areas as per the BPM hierarchy, (c) Finally, QBRA makes recommendations based upon the search flow (i.e., BPM selections) including references to explore further.}
	\label{fig:prototype}
\end{figure*}

A web based tool (see Figure~\ref{fig:prototype}) has been designed to realize the QBRA design presented here. Tool guides its users to navigate (proprietary) BPM and finally connects that to the underlying computational problems for which recommendation is shown to the user as final output. 

Figure~\ref{fig:prototype} presents selected screens of the tool starting with selection of the industry to underlying problems application might be focused upon and finally recommendation by the tool including further QC Resources to explore. 

During initial interactions involving demonstrations of prototype tool to QC software developers, tool idea received quite positive response, especially, in bridging the gap for business executives and giving them an informed glimpse of what the current-state-of-the-art in QC literature in respect to their focused domain is and what will it take to start building initial prototypes for explorations.


\section{Conclusion and Future Work}
Unless current limitations of QC technology are successfully overcome, only a limited class of problems if solved using it may offer significant business advantage over existing classical approaches~\cite{8}, \cite{preskill2018quantum}, \cite{corcoles2019challenges}, \cite{monroe2019us}. Thus, not all applications having business potential would benefit by being implemented as quantum programs. Often the ones that can be solved today using QC assume adiabatic hardware (e.g., D-Wave system~\cite{5}) with commercial implementations still sometime away. Many of the problems optimally solvable on QCs can also be solved on classical computing environments albeit with reduced performance. To make such considerations clearer to business professionals, in this work we presented an approach to augment users while deciding to invest into quantum software development as a plausible future option in their application context. 

There are directions where the approach can be refined further including detailed pilot studies to determine the depth at which business users find guided exploration helping them to understand QC technology better in the context of their business. Currently it is driven by the business process model of the industry and associated list of known computational problems where QC (or CC) is considered to offer relative advantages. Another direction is to augment problem database and corresponding text corpus associated with each problem by automatically extracting accessible knowledge sources on the web or internal knowledge repositories of the organization. 

Though a bit challenging but automatically harvesting the QBRA problem database as and when new quantum algorithms or applications are reported in the literature and inferring associations of these with business processes could be an interesting direction for future research. Currently QBRA problem database is manually curated, in particular, while associating leaf nodes in the business process hierarchy with the computational problems and associated quantum/classical algorithms. As the field of algorithm design is constantly evolving, so does this database also requires ongoing updates, where any degree of automated assistance would help reducing manual dependence and in turn deployment cost. 

For problem scenarios where quantum processors would provide differentiating advantage over classical one, integration with approaches like one presented in~\cite{Salm2020NISQAnalyzer} towards selection of plausible implementation of quantum solution and suitable quantum processor to execute selected implementation would take current approach to next level of application in practice.   
  
\bibliographystyle{abbrv}
\bibliography{qc}

\end{document}